# Substrate Stiffness Changes Cell Rolling and Adhesion over L-selectin Coated Surface in a Viscous Shear Flow

Maryam Bagheri[1*], Mohammadreza Azmoodeh[2]

**Authors' information**
[1]D. Email: mbaghe3@uic.edu
[2]D. E-mail: m.azmoode@ut.ac.ir
*Corresponding author

**Abstract:** Understanding the rolling and adhesion behavior of a cell on the vascular surface under viscous shear flow is important to better understand many biological processes. One of the important examples is the adhesion of the leukocytes onto sites of an inflammation. Recently, researchers have started to investigate the effects of surface rigidity on the cell rolling and adhesion. Inspired by recent studies, we employed Adhesive dynamics to investigate effects of substrate rigidity on a rolling cell in a hydrodynamic flow. The vascular surface is modeled as an elastic surface with different Young's modulus's, coated with L-selectin molecules, which can form bonds with PSGL-1 ligands on the cell. The results illustrate that the stiffer substrate helps to capture cells. The results of this study help to better understand the adhesion of a cell by L-selectin coated surface.

## 1. Introduction

One of the fundamental processes in any biological system is the adhesion of a cell to a surface or another cell. The process of adhesion of a cell to a surface is time dependent and dynamic, which starts with forming bonds between ligands on the cell and receptors on the substrate [1-4]. The best example, probably, is the adhesion of leukocytes to an injured tissue of an inflammation site. During this process the rolling leukocytes first form tethers with the activated endothelium and this process is mediated by the presence of selectins [5]. The common selectins that are found on the endothelium cells are P-, L-, and E-selectin [1].

L-selectin is a type of glycoprotein that is usually found on the microvilli of the many leukocytes [6, 7]. L-selectins are known to be associated with many diseases like HIV, diabetes type II, lymphoma, sepsis, and stoke [8-10]. The main two functions of L-selectins are 1. Guiding lymphocytes to lymph nodes and 2. directing leukocytes to the sites of inflammation [11, 12]. PSGL-1 is a mucin-like glycoprotein, which is a common ligand for P-, E-, and L-selectins [1]. One of the frequent ligands on the cell are PSGL-1, which can form bonds with L-selectin coated surface under viscous shear flow condition [5, 13]. The bond formation and dissociation between the ligand and the receptors are attributed to the bond formation ($k_f$) and rupture ($k_r$) rates, as well as, the developing force in the tether. As shown by [13, 14], the cell adhesion dynamics is primarily dominated by the physical chemistry of the ligand and the receptors, however, as illustrated by [2, 16], the elasticity of the underlying substrate can affect the rolling and adhesion of the cells in many circumstances. Many recent experimental studies investigated the particle interactions with each other or the surface either in Newtonian or non-Newtonian shear flow [17-23]. Indeed, a recent study [16] has shown that the rigidity of the surface can change the adhesion behavior of the cells rolling over an E-selectin coated surface.

In recent computational studies [1, 2], Adhesive Dynamics (AD) simulation [14] was modified in order to consider the effects of substrate elasticity on the rolling and adhesion of the cells to the substrate in a viscous shear flow. Moshaei et al. [2] employed a linear elastic model [24] to add the surface elasticity to the AD simulation. In this model, the authors used two-spring configuration (as shown in figure 2) to model the bond between the cell and the substrate. They further investigated the effects of the substrate rigidity on the cell adhesion to the E- and P-selectin coated surface and they showed that their model can reproduce the experimental results presented by [16].

In this paper we used Moshaei's model [1, 2] to investigate the effect of surface elasticity on the rolling and the adhesion of the cell with PSGL-1 ligands to the substrate coated with L-selectin.

## 2. Methodology

In this section we the methodology, which is used in this paper based on the modification that Moshaei et al. [1, 2] performed on AD model. Hammer and his coworker outlined more details about AD computational modeling in their earlier works [14, 15]. The cell is considered to be a rigid sphere in a viscous shear flow with viscosity $\mu$ and shear rate $\dot{\gamma}$. The cell moves near soft surface, which is modeled based on a linear elastic model with Young's modulus $E$ and Poisson's ratio $\nu = 0.5$. Following Moshaei et al. [2] the surface is coated with L-selectin receptors and is assumed to be incompressible.

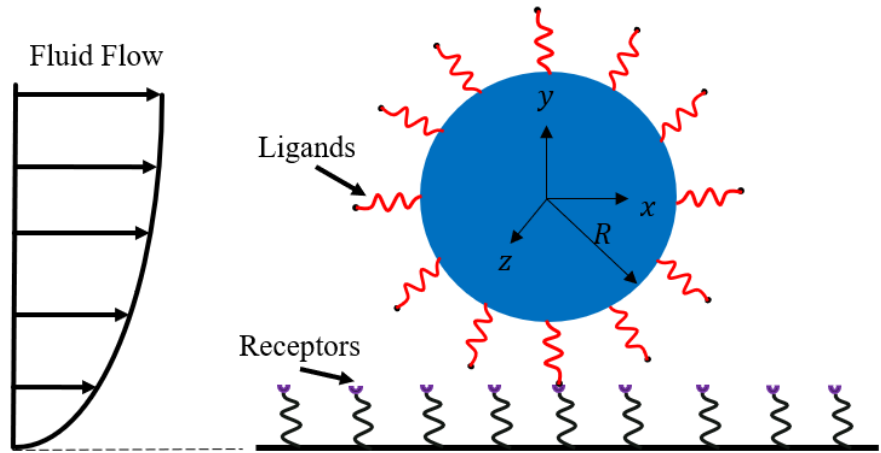

*Figure 1. Schematic of the cell covered with ligand rolling along a substrate coated with receptors in viscous shear flow*

### 2.1. Bond Formation and Rupture

Bond formation and rupture between the ligands on the cell and the receptors on the substrate is a stochastic process, which depends on forward reaction rate and reverse reaction rates, $k_f$ and $k_r$, respectively. The bond rupture rate depends on the force, which develops in the bond and Bell [25] formulated this as

$$k_r = k_r^0 \exp\left(\frac{\gamma f}{K_B T}\right)$$

Here, $k_r^0$ is intrinsic reverse rate, $\gamma$ is the reactive compliance, $f$ is the bond force, and $K_BT$ is the thermal energy.

The rate of bond formation follows detailed balance and based on Bell's model [25] is expressed as

$$k_f = k_f^0 \exp\left(\frac{k_b|\Delta - l_b^0|(\gamma - 0.5|\Delta - l_b^0|)}{2K_BT}\right)$$

Here, $k_f^0$ is the intrinsic forward rate, $\Delta$ is the separation distance between the ligands and the receptors, $l_b^0$ is the equilibrium bond length, and $k_b$ is the bond stiffness.

Considering $l_b$ as the bond length after deformation, the force in the bond is

$$f = k_b(l_b - l_b^0)$$

As illustrated in the figure 2 the total stiffness of the bond and substrate can be expressed as

$$\bar{k} = \frac{k_b k_s}{k_b + k_s}$$

Where, $k_s$ is the equivalent surface stiffness and can be formulated based on Kendall [24] model as

$$k_s \approx \frac{E l_b^0}{1 - \nu^2}$$

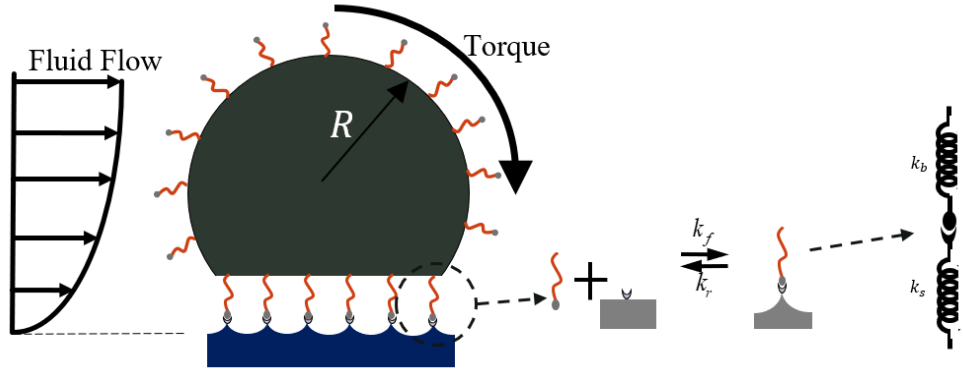

*Figure 2. Schematic of the Moshaei's model [] for substrate elasticity*

## 2.2. Equilibrium Condition

Assuming the cell is small enough and moves in a creeping flow, it is acceptable to neglect the effects of inertia and write the equilibrium condition for the cell as

$$\boldsymbol{F}^b + \boldsymbol{F}^s + \boldsymbol{F}^h = 0$$

Where, $\boldsymbol{F}^b$ is the force in the all bonds, $\boldsymbol{F}^s$ is the resultant non-specific forces, and $\boldsymbol{F}^h$ is the hydrodynamic force.

Non-specific forces in this simulation are, the gravity ($F_g^s$), the Van der Walls ($F_v^s$), the electrostatic ($F_e^s$), and the steric stabilization forces ($F_s^s$) and are defined below

$$F_g^s = \frac{4}{3}\pi(\rho_c - \rho_f)R^3 g$$

Here, $\rho_c$ is the cell density, $\rho_f$ is the fluid density, and R is the cell radius.

$$F_e^s = 2\pi R \frac{\rho^2}{2\epsilon\kappa^3}(e^{\kappa L} - 1)^2 e^{-\kappa h} \qquad d > 2L$$

$$F_e^s = 2\pi R \left(\frac{\rho^2}{2\epsilon\kappa^2}(4L - d2) + \frac{\rho^2}{2\epsilon\kappa^3}(1 - 2e^{\kappa L})^2 e^{-\kappa h} + \frac{\rho^2 e^{-\kappa h}}{2\epsilon\kappa^3} e^{-2\kappa L}\right) \qquad d < 2L$$

Here, $\rho$ is the charge density, $\kappa$ is Debye-Huckel length, $\epsilon$ is the dielectric constant, $d$ is the gap distance between the cell and the substrate, and $L$ is the cell coat thickness.

$$F_s^s = 0 \quad d > 2L$$
$$F_s^s = 2\pi R \frac{\lambda(2L - d)}{L^2} \quad d < 2L$$

Here, $\lambda$ is the steric constant.

$$F_v^s = -\frac{A_h R}{6}\left(\frac{1}{d^2} + \frac{1}{(d+2H)^2} - \frac{2}{(d+H)^2}\right)$$

In the Van der Walls force, $A_h$ is the Hamaker constant and $H$ is a constant on the order of $d$.

Goldman et al. [26, 27] formulated shear force $F^h$ and shear couple $C^h$ as shown below
$$F^h = 6\pi\mu R(R + d)\dot{\gamma}F^{h*}$$

$$C^h = 4\pi\mu R^3 \dot{\gamma} C^{h*}$$

Here, $F^{h*}$ and $C^{h*}$ are given by [26, 27] as a function of the distance between the cell and the substrate.

Mobility matrix can be used to re-write the equilibrium condition as
$$\boldsymbol{U} = \boldsymbol{M}.\boldsymbol{F}$$
Where, $\boldsymbol{U}$ the velocity vector, $\boldsymbol{M}$ is the mobility matrix, and $\boldsymbol{F}$ the force vector. Following [14], the velocity and force vectors are
$$\boldsymbol{U} = (V_x, V_y, V_z, \Omega_x, \Omega_y, \Omega_z)$$
$$\boldsymbol{F} = (F_x^b + F^h, F_y^b + F^s, F_z^b, C_x^b, C_y^b, C_z^b + C^h)$$
Where, $V$ is the linear velocity and $\Omega$ is the angular velocity. The mobility matrix is a $6 \times 6$ matrix and its components are

$M^{11} = M^{33} = \frac{T_r}{6\pi\mu RD}$, $M^{22} = \frac{1}{6\pi\mu R\lambda_g}$, $M^{34} = -M^{16} = \frac{T_t}{6\pi\mu R^2 D}$, $M^{43} = -M^{61} = \frac{T_r}{8\pi\mu R^2 D}$, $M^{44} = M^{66} = \frac{T_t}{8\pi\mu R^3 D}$, $M^{55} = \frac{1}{8\pi\mu R^3 X_g}$

Where, $T_t, T_r, F_t,$ and $F_r$ are defined in [26, 27] and values for $\lambda_g$ and $X_g$ are defined by [28]. $D$ is
$$D = T_t F_r - F_t T_r$$

### 2.3. Monte Carlo Simulation

Following the AD computational procedure, we adopted Monte Carlo simulation to update the velocity of the cell at each time step. At each time step all the exerted forces on the cell are calculated by using the formulation above and then using the probability functions the Monte Carlo algorithm updates the number of the bonds between the cell and the surface by determining if new bonds are formed or the existing bonds are ruptured. Following [14] we used the bond formation probability $P_f$ and bond rupture probability $P_r$ as defined below
$$P_f = 1 - \exp(-k_f \Delta t)$$

$$P_r = 1 - \exp(-k_r \Delta t)$$
Considering computational efficiency, we considered $\Delta t = 10^{-6}$ s.

The values that used for the parameters in the non-specific forces are presented in table 1 and the values for the AD simulation are presented in table 2.

Table 1. Estimates for the constant values appeared in non-specific interactions [1, 2]

| Constant | Value | Unit | Definition |
|---|---|---|---|
| $\rho$ | 0.05 | $g/cm^3$ | Charge density |
| $\rho_c$ | 1 | $g/cm^3$ | Cell density |
| $\rho_f$ | 1.05 | $g/cm^3$ | Fluid density |
| $H$ | 70 | $A^0$ | Surface thickness |
| $\epsilon$ | $7.8 \times 10^{-19}$ | $C^2/dyn.cm^2$ | Dielectric constant |
| $L$ | 10 | $nm$ | Cell coat thickness |
| $\kappa$ | 0.125 | $A^0$ | Debye-Huckel length |
| $A_h$ | $5 \times 10^{-21}$ | $J$ | Hamaker constant |
| $\lambda$ | $2.5 \times 10^{-6}$ | $dyn$ | Steric constant |

Table 2. Values for AD modeling parameters [1, 15, 29]

| Constant | Value | Unit | Definition |
|---|---|---|---|
| $R$ | 5 | $\mu m$ | Cell radius |
| $k_f^0$ | 84 | $s^{-1}$ | Intrinsic forward rate |
| $k_r^0$ | 12.7 | $s^{-1}$ | Intrinsic reverse rate |
| $\dot{\gamma}$ | 100 | $s^{-1}$ | Shear rate |
| $k_b$ | 1 | $dyn/cm$ | Bond stiffness |
| $\mu$ | 1 | $g/cm.s$ | Viscosity |
| $l_b^0$ | 25 | $nm$ | Equilibrium bond length |
| $N_l$ | 3500 | – | Ligand density |
| $\Delta t$ | $10^{-6}$ | $s$ | Time step |
| $\gamma$ | 0.15 | $A^0$ | Reactive compliance |

3. **Results and Discussions**

In this section we present the results of our simulation. The AD algorithm is employed as discussed in the previous section to simulate rolling and adhesion of a cell with PSGL-1 ligand to a surface coated with L-selectin. The substrate Young's modulus used for this simulation is between 1 to 100 kPa, with 5 kPa for very compliant surface, 20 kPa, and 80 kPa for a stiffer surface. As stated in the table 2 the shear rate is $100\ s^{-1}$ and the starting gap between the cell and the substrate is $30\ nm$.

Figure 3-5 show how the velocity of the cell changes over different substrate elasticity during 30 seconds of the simulation. The average velocity of the cell over the surface with $E = 5\ kPa$ is $1.73\ m/s$, for the surface with $E = 20\ kPa$ is $1.58\ m/s$, and $1.12\ m/s$ for the surface with $E = 80\ kPA$.

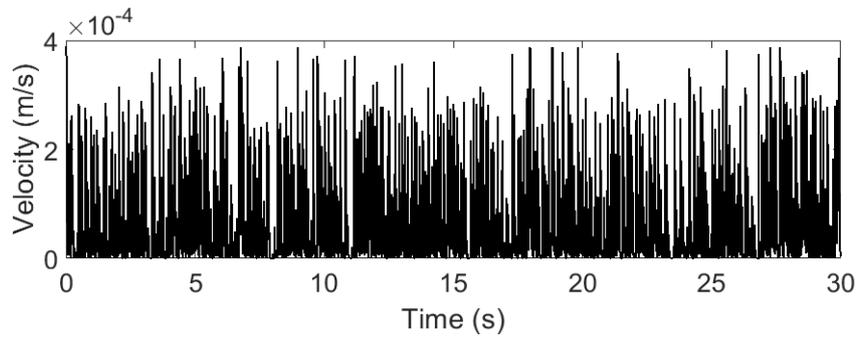
*Figure 3. Instantaneous velocity of a cell for 30 seconds simulation over substrate with E=5 kPa*

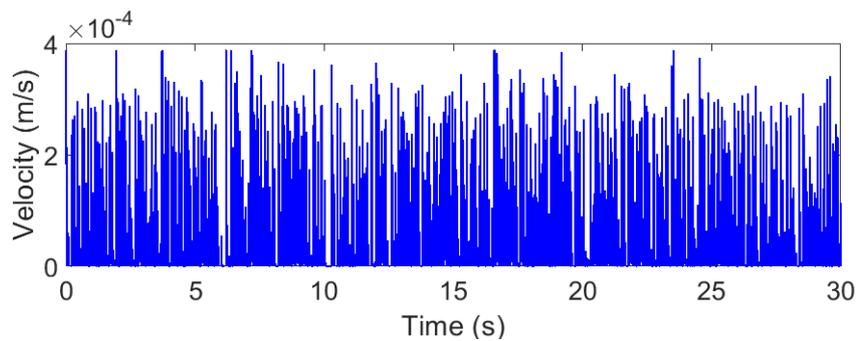
*Figure 4. Instantaneous velocity of a cell for 30 seconds simulation over substrate with E=20 kPa*

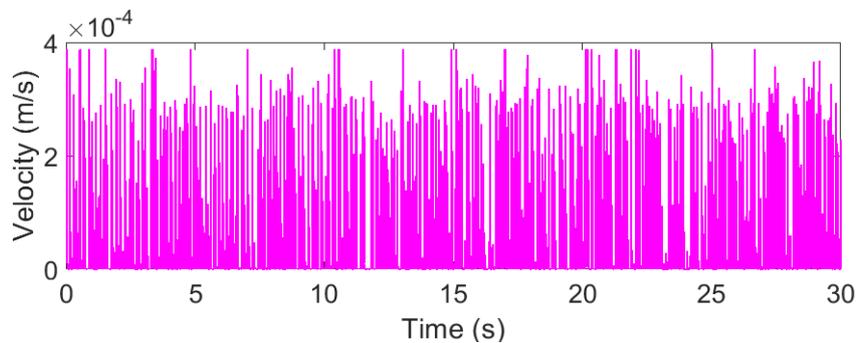
*Figure 5. Instantaneous velocity of a cell for 30 seconds simulation over substrate with E=80 kPa*

Figure 6 shows the number of the bonds at each time step. We plotted all three results together to show that the number of the bonds are close and follows a similar trend when the surface elasticity is changed. The average number of bonds is 7 for $E = 5\ kPa$, 6 for $E = 20\ kPa$, and 4 for $E = 80\ kPa$. Although the average number is higher for the softer surface, but the fluctuations are higher, this means that when the surface is stiffer it can resist against rupture of the formed bonds better than a softer surface. This means that the formed bond between the cell and the substrate can survive longer and gradually help to capture the cell, while on the softer surface, the cell may escape and roll with the flow again more easily.

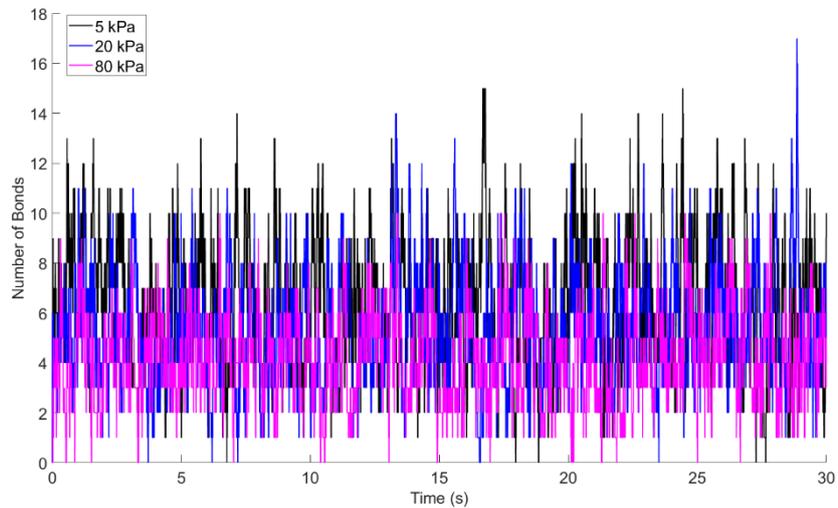

*Figure 6. Total number of bonds between a cell and a substrate for 30 seconds simulation over substrate with different elasticity.*

Figure 7 shows the travelled distance of the cell over surfaces with different elasticities. As it could be predicted base on [2], the stiffer substrate helps to capture the cell. It is important to notice that the differences between these three simulations are not big, which is due to the intrinsic forward and backward reaction rate. As stated in [2, 16], surface stiffness is not the only parameter in adhesion of a cell to a substrate. Many different parameters alter this biological process and the rates of bond formation and dissociation are two examples. So, as it can be seen in figure 7, compared to E- and P-selectin coated surfaces, cells can travel more over L-selectin coated surface in general. However, this figure shows that the stiffer surface can still help the adhesion of the cells compared to the softer surfaces.

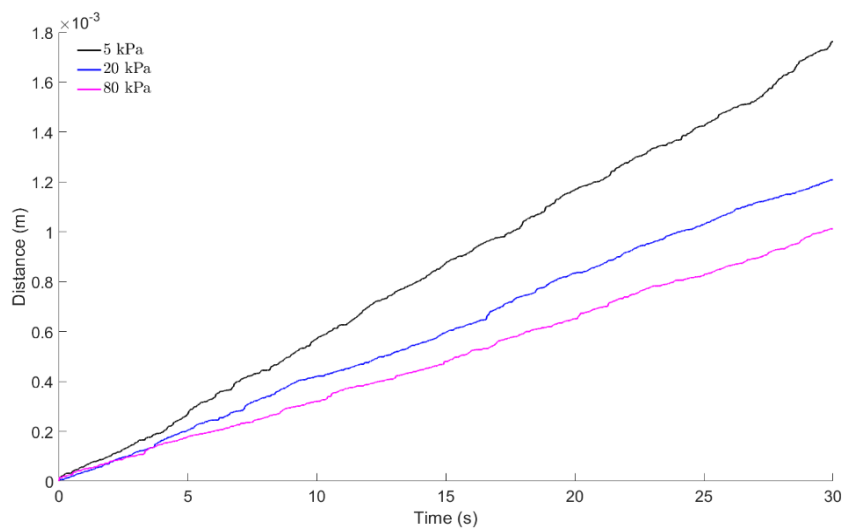

*Figure 7. Travelled distance of a cell over substrates with different elasticities.*

# 4. Conclusion

In this study we investigated the effect of substrate rigidity on rolling and adhesion of a cell over a surface coated with L-selectin. We employed AD simulation [14] and adopted the method introduce by Moshaei et al. [1-3] to consider the effects of substrate elasticity. For the bond formation and dissociation parameters we used data provided by [21]. Our simulations were completed for 3 different Young's modulus's, 5, 20, and 80 kPa to see how the shift in the elasticity would alter the rolling and adhesion behaviors.

As illustrated in the results, the stiffer substrate can help to capture the cells, however, it is important to mention that our results are for 30 seconds of the simulation and the results do not show that the cell is stopping. This is consistent with what has been predicted in the state diagram in [2]. As previously stated by [2, 16], substrate elasticity tends to have different effects in different set-ups. For the scope of our investigation, we can clearly see the effects of different Young modulus's, however, since the intrinsic dissociation rate in large compared to the previous studies [2, 16], the cell can travel with less resistive bond forces between PSGL-1 and L-selectin.